\documentclass[12pt]{revtex4}

\newcommand{\plb}[2]{{\em Phys. Lett.}              {\bf #1B}, #2 }
\newcommand{\npb}[2]{{\em Nucl. Phys.}              {\bf B#1}, #2 }
\newcommand{\npp}[2]{{\em Nucl. Phys. Proc. Suppl.} {\bf  #1}, #2 }
\newcommand{\pr }[2]{{\em Phys. Rep.}               {\bf  #1}, #2 }
\newcommand{\prt}[2]{{\em Phys. Rev.}               {\bf D#1}, #2 }
\newcommand{\pru}[2]{{\em Phys. Rev. Lett.}         {\bf  #1}, #2 }
\newcommand{\zpc}[2]{{\em Z. Phys.}                 {\bf C#1}, #2 }

\newcommand{\epc}[2]{{\em Eur. Phys. J.}            {\bf C#1}, #2 }
\newcommand{\ijm}[2]{{\em Int. J. Mod. Phys.}       {\bf A#1}, #2 }
\newcommand{\con}[2]{                               {\bf  #1}, #2 }
\newcommand{\etal}{{\em et al.}}
\newcommand{\ibid}{{\em ibid.}}

\newcommand{\aspi}{{\hat\alpha_s\over \pi}}
\newcommand{\aspis}{{\hat\alpha_s^2 \over \pi^2}}
\newcommand{\five}{\hspace{5pt}}

\newcommand{\be}{\begin{equation}}
\newcommand{\ee}{\end{equation}}
\newcommand{\ben}{\begin{eqnarray}}
\newcommand{\een}{\end{eqnarray}}
\newcommand{\ba}{\begin{array}}
\newcommand{\ea}{\end{array}}

\def\msbar{\mbox{{\footnotesize{$\overline{\rm MS}$}}} }

\newcommand{\lsim}{\buildrel < \over {_\sim}}

\def\al2{\frac{\alpha^2}{\pi^2}}

\begin{document}

\title{
QCD Sum Rules and the Determination of Fundamental Parameters\footnote{
Presented in the 8th Accelerator and Particle Physics Institute (APPI 2003),
Appi, Iwate Ken, Japan, Feb 25-28 2003.}}
\author{Jens Erler$^{a}$ and Mingxing Luo$^{b,c}$}
\affiliation{
$^{a}$Instituto de F\'\i sica, Universidad Nacional Aut\'onoma de M\'exico, 
Apdo.\ Postal 20-364, 01000 M\'exico D.F., M\'exico \\
$^{b}$Zhejiang Institute of Modern Physics, Department of Physics\\
Zhejiang University, Hangzhou, Zhejiang 310027, P.R. China\\
$^{c}$Theory Division, IPNS, High Energy Accelerator Research Organization, 
1-1 Oho, Tsukuba-shi, Ibaraki-ken, 305-0801, Japan}

\date{\today}

\begin{abstract}
We present a new QCD sum rule with high sensitivity to the continuum regions of
charm and bottom quark pair production.  Combining this sum rule with existing 
ones yields very stable results for the \msbar quark masses, 
$\hat{m}_c (\hat{m}_c)$ and $\hat{m}_b (\hat{m}_b)$. 
Comparison of our approach with experimental data allows for a robust theoretical error 
estimate. 
We have also provided a new evaluation of the lifetime of 
the $\tau$ lepton, $\tau_\tau$, serving as a strong constraint on $\alpha_s$.  
\end{abstract}

\maketitle

\section{Introduction}
The determination of the fundamental Standard Model (SM) parameters is 
important in its own right. It provides a test of the SM when results from 
various sources are compared, which can foster our understanding 
of SM dynamics (such as strong QCD effects).
This may also lead to hints of new physics beyond the SM, 
when precise values of the SM parameters 
are compared against the predictions of more fundamental theories.
For example, gauge couplings do not unify within the SM.
This gives extra evidence against simple grand unification theories 
(GUTs) such as $SU(5)$ without supersymmetry,
in addition to the non-observation of proton decay.
On the other hand, gauge couplings seem to unify
at a scale $\sim 2\times 10^{16}$~GeV in the minimal
supersymmetric standard model, which can be interpreted
as a hint for supersymmetry as well as GUTs \cite{GUT}.
Most GUTs \cite{Langacker:1980js} also predict the mass ratio $m_b/m_\tau$.

It is generally difficult to obtain reliable information on quark masses.  
The Particle Data Group~\cite{Hagiwara:2002} lists only ranges for their 
values, indicating a lack of confidence in methods used 
to evaluate them.  Indeed, $\alpha_s$ is quite large at the mass scales of 
the bottom and charm quarks, questioning the convergence of perturbative QCD 
(PQCD).  Furthermore, non-perturbative effects governed by 
the scale $\Lambda_{\rm QCD} \sim 0.5$~GeV could be large, thus potentially
compromising the validity of perturbative calculations.  
Two types of conditions are known to improve 
the situation:  high energy or inclusiveness.  As an example for the former, 
$\alpha_s$ and $m_b$ can be determined at LEP energies using PQCD. This yields
$\alpha_s(M_Z) = 0.1200\pm 0.0028$~\cite{Erler:sa} with very little theoretical
uncertainty. But $b(c)$ quark effects are small, so that  
$\hat{m}_b(M_Z)= 2.67\pm 0.50$~GeV~\cite{Abreu:1997ey} is not well constrained.

In a recent work \cite{plb}, we computed $\alpha_s$ from $\tau_\tau$, by 
definition 
an inclusive quantity and known to be quite insensitive to effects from 
non-perturbative QCD (NPQCD)~\cite{Braaten:1991qm}.  Likewise, we used
a set of inclusive QCD sum rules to derive values for $\hat{m}_c (\hat{m}_c)$ 
and $\hat{m}_b (\hat{m}_b)$.  One of these sum rules is new, and its use 
together with existing ones~\cite{Novikov:et,Shifman:bx} proves to be 
a powerful tool to constrain the continuum region of quark pair production.  
This will be particularly helpful for the case of the $b$ quark
for which precise measurements of $R(s)$ (the inclusive hadronic cross section 
normalized to the leptonic point cross section) or of $R_b(s)$ 
(exclusive cross section for $b\bar{b}$ pairs) are unavailable. 

In section 2, we determine the heavy quark masses
based upon the new sum rule in addition to known ones.
We point out a puzzling discrepancy between theory and the recent BES data.
The BES data seem to be lower than theoretical predictions 
by $30 \%$ consistently across the moments.
In section 3, we compute the $\tau$ lifetime.
In section 4, we summarize our results.

\section{Sum rules and heavy quark masses}
On the basis of an unsubtracted dispersion relation (UDR) it was shown 
in Ref.~\cite{Erler:1998sy} that knowledge of $m_c$, $m_b$, and $\alpha_s$ 
is sufficient to compute the charm and bottom quark contributions to 
the QED coupling $\alpha (\sqrt{t} = M_Z)$.
Conversely, 
comparison of this UDR with the more traditional approaches using 
a subtracted dispersion relation (SDR) offers information on $m_c$ and 
$m_b$.  The resulting equation relates an inclusive integrated cross 
section to a difference of vacuum polarization tensors, {\it viz.}
\be
   12\pi^2 \left[ \hat\Pi_q (0) - \hat\Pi_q (-t) \right]=
   t \int_{4 m_q^2}^\infty {{\rm d} s\over s} {R_q(s)\over s + t}.
\label{sumrulet}
\ee
Eq.~(\ref{sumrulet}) defines a continuous set of sum rules parametrized 
by $t$, where the limit $t \rightarrow 0$ coincides with the first 
moment of $\Pi_q (t)$. Similarly, 
for each higher moment, ${\cal M}_n$, 
one has~\cite{Novikov:et,Shifman:bx,Voloshin:1995sf,Eidemuller:2000rc,Kuhn:2001dm}. 
\be
   \left.{12\pi^2\over n !} {d^n\over d t^n} \Pi_q(t) \right|_{t=0}
   = \int_{4 m_q^2}^\infty {{\rm d} s\over s^{n+1}} R_q(s).
\label{sumrulen}
\ee
We now take the opposite limit in Eq.~(\ref{sumrulet}), $t\rightarrow\infty$,
and regularize the divergent expression,
\be
   {R_q (s)\over 3 Q_q^2} \longrightarrow {R_q (s)\over 3 Q_q^2} - 
   \lambda^q_1 (s) \equiv {R_q (s)\over 3 Q_q^2} - 1 -  
   {\alpha_s (\sqrt s) \over \pi}
\label{regular}
\ee
\vspace{-12pt}
$$   - \left[ {\alpha_s (\sqrt s) \over \pi} \right]^2
       \left[ {365\over 24} - 11 \zeta(3)
            + n_q \left( {2\over 3} \zeta(3) - {11\over 12} \right) \right].
$$
$Q_q$ and $n_q$ are the quark charge and the number of active flavors.  
Using expressions derived in Refs.~\cite{Chetyrkin:1996cf,Chetyrkin:1997un}
and taking the limit $t \rightarrow \infty$, 
the sum rule~(\ref{sumrulet}) becomes:
\ben
   \sum\limits_{\rm resonances} {3\pi\Gamma^e_R\over Q_q^2 M_R 
   \hat\alpha^2 (M_R)} + \int\limits_{4 M^2}^\infty {{\rm d} s\over s}  
   {R_q^{\rm cont}\over 3 Q_q^2} - \int\limits_{\hat{m}_q^2}^\infty 
   {{\rm d} s\over s} \lambda^q_1 (s) 
   = - {5\over 3} + \aspi \left[ 4 \zeta(3) - {7\over 2} \right] \nonumber \\
 + \aspis  \left[ {11\over 4} \zeta(2) + {2429\over 48} \zeta(3) - \right. 
   \left. {25\over 3} \zeta(5) - {2543\over 48} + n_q \left( {677\over 216} -
   {\zeta(2)\over 6} - {19\over 9} \zeta(3) \right) \right].
\label{sumrule0}
\een
Here, $M_R$ and $\Gamma^e_R$ are the mass and the electronic partial width of 
resonance $R$, and $R_q^{\rm cont}$ denote the continuum regions integrated 
from $M = M_{B^\pm}$ for $b$ and $M = M_{D^0}$ for $c$.
The regularization~(\ref{regular}) together with the scale 
choices $\hat{m}_q = \hat{m}_q (\hat{m}_q)$ and 
$\hat\alpha_s = \hat\alpha_s (\hat{m}_q)$ eliminates (resums) all logarithmic 
terms in Eq.~(\ref{sumrule0}). Unlike in any of the sum 
rules~(\ref{sumrulen}), $R_q^{\rm cont}$ appears unsuppressed in 
Eq.~(\ref{sumrule0}) so that $\hat{m}_q$ varies exponentially with 
the experimental information on the resonances.
We will use  Eq.~(\ref{sumrule0}) 
to constrain the continuum region and work with the following {\em ansatz\/}:
\ben
   {R_q^{\rm cont} (s)\over 3 Q^2_q} & = & \lambda^q_1 (s) 
   \sqrt{1 - {4\, \hat{m}_q^2 (2 M) \over s^\prime}} 
   \left[ 1 + \lambda^q_3 {2\, \hat{m}_q^2(2 M) \over s^\prime} \right] \nonumber \\
& \approx &
   \lambda^q_1 (4 M^2) \sqrt{1 - {4\, \hat{m}_q^2 \over s^\prime}} \left[ 1 + 
   \lambda^q_3 {2\, \hat{m}_q^2 \over s^\prime} \right] - \aspi
   {\lambda_2^q (s)\over 1 + \lambda_2^q(s)},
\label{ansatz}
\een
where now $\hat\alpha_s = \hat\alpha_s (2 M)$, 
$s^\prime \equiv s + 4 (\hat{m}_q^2 (2 M) - M^2)$, and
$$
  \lambda_2^q(s) = \aspi \beta_0 \ln {s\over 4 M^2} = {\hat\alpha_s (2 M)\over 
  \pi} \left( {11\over 4} - {n_q\over 6} \right) \ln {s\over 4 M^2}.
$$ 
We will use the form in the second line (applying it to all moments) of 
Eq.~(\ref{ansatz}) with the corresponding change in the regularization in 
Eq.~(\ref{sumrule0}). This keeps only the leading logarithms resumed but 
allows for an analytical integration.  
Eq.~(\ref{ansatz}) coincides asymptotically with the predictions 
of PQCD for massless quarks and interpolates smoothly between the vanishing 
phase space at the pseudo-scalar threshold and the strong onset of fermion pair 
production.
Unlike when PQCD is 
applied to $R(s)$ directly and relatively close to the resonance region, we 
minimize the exposure to local quark-hadron duality violations by using QCD 
{\em inclusively\/} and by merely requiring stable results across the moments. 
No claim is being made about the {\em local shape\/} of $R_q$ --- we only need 
theoretical information about {\em global averages\/}.
It should be pointed out that the new sum rule (4) and the choice of 
{\em ansatz\/}~(\ref{ansatz}) are
logically independent ingredients. An explicit {\em ansatz\/} facilitates the
discussion, but our results are essentially independent of the shape of the continuum.

We use the narrow resonance data~\cite{Hagiwara:2002},
$J/\Psi$, $\Psi(2S)$ for the $c$ quark and $\Upsilon(1S)$, $\Upsilon(2S)$, $\Upsilon(3S)$ 
for the $b$ quark,
as the only experimental input.  The wider resonances in
the continuum region are assumed to be accounted for by our 
{\em ansatz\/}~(\ref{ansatz}) because (i) they decay almost exclusively into 
flavored hadrons; (ii) they interfere with the non-resonating part of 
the continuum rendering a common treatment virtually impossible; 
(iii) the $\delta$-function approximation (which is perfect for the narrow
resonances) becomes successively worse; (iv) the philosophy of our 
{\em ansatz\/} supposes that it averages over local cross-section fluctuations;
and (v) we wish to compare Eq.~(\ref{ansatz}) directly to experimental data on
the charm continuum region such as from Beijing~\cite{Bai:2001ct}.
\begin{table} 
\begin{tabular}{|c|c|r|r|r|}
$n$ & resonances & continuum & total & theory \hspace{8pt} \\
\hline
0 & 1.16 (6) &$-3.03 \pm 0.37$ &$-1.86 \pm 0.37$ & {\bf input} (\five 4) \\
1 & 1.12 (6) & $1.04 \pm 0.14$ & $2.16 \pm 0.16$ & 2.19        (\five 6) \\
2 & 1.10 (7) & $0.37 \pm 0.07$ & $1.47 \pm 0.10$ & 1.49        (\five 9) \\
3 & 1.10 (7) & $0.17 \pm 0.04$ & $1.27 \pm 0.08$ & 1.26             (14) \\
4 & 1.11 (7) & $0.09 \pm 0.02$ & $1.20 \pm 0.08$ & 1.16             (20) \\
5 & 1.13 (7) & $0.05 \pm 0.01$ & $1.18 \pm 0.08$ & 1.10             (31) \\
\hline
0 & 1.17 (5) &$-52.44\pm 1.24$ &$-51.27\pm 1.24$ & {\bf input} (\five 2) \\
1 & 1.24 (5) & $3.12 \pm 0.53$ & $4.36 \pm 0.54$ & 4.51        (\five 2) \\
2 & 1.31 (5) & $1.33 \pm 0.30$ & $2.64 \pm 0.31$ & 2.79        (\five 3) \\
3 & 1.40 (5) & $0.75 \pm 0.19$ & $2.15 \pm 0.20$ & 2.27        (\five 5) \\
4 & 1.50 (5) & $0.48 \pm 0.13$ & $1.98 \pm 0.14$ & 2.06        (\five 7) \\
5 & 1.61 (5) & $0.33 \pm 0.10$ & $1.94 \pm 0.11$ & 1.99             (10) \\
6 & 1.74 (6) & $0.23 \pm 0.07$ & $1.98 \pm 0.09$ & 1.98             (14) \\
7 & 1.89 (6) & $0.17 \pm 0.05$ & $2.06 \pm 0.08$ & 2.03             (19) \\
\end{tabular} 
\caption[]{Results for the lowest moments, ${\cal M}_n$, defined in 
Eq.~(\ref{sumrulet}) for $n = 0$ ($t \rightarrow \infty$) and 
Eq.~(\ref{sumrulen}) for $n \geq 1$.  The upper (lower) half of the Table 
corresponds to the charm (bottom) quark.  Each moment has been multiplied by 
$10^n\mbox{GeV}^{2n}$ ($10^{2n+1}\mbox{GeV}^{2n}$). The continuum error is 
from $\Delta\lambda_3^{b,c} = \pm 1.47$. The last column shows the theoretical 
prediction for $\hat{m}_c (\hat{m}_c) = 1.289$~GeV, 
$\hat{m}_b (\hat{m}_b) = 4.207$~GeV, and $\alpha_s (M_Z) = 0.1211$, where 
the uncertainty is our estimate for the truncation error (see text).}
\label{moments}
\end{table}
The narrow resonance  contribution to the various moments is shown in 
the second column of Table~\ref{moments}.
The 3rd 
column gives the continuum contribution, and the 4th column shows the totals 
to be compared with the theoretical moments in the last column, {\em viz.}
\be
   {\cal M}_n^{\rm theory} = {9\over 4} Q_q^2 
   \left( {1\over 2\hat{m}_q(\hat{m}_q)} \right)^{2n} \bar{C}_n.
\label{theory}
\ee
The $\bar{C}_n$ are known up to ${\cal O}(\alpha_s^2)$ and taken from
Refs.~\cite{Chetyrkin:1996cf,Chetyrkin:1997mb} where they were computed
for arbitrary renormalization scale $\mu$.  It seems appropriate to choose 
$\mu = \hat{m}_q (\hat{m}_q)$, eliminating all logarithmic terms as there is 
only one scale in the problem.  Indeed, the authors of Ref.~\cite{Kuhn:2001dm},
who have chosen $\mu = 3$~(10)~GeV for the charm (bottom) quark and then 
evolved to $\mu = \hat{m}_q$, report a variation over the first 5~(7)~moments 
of 122~(312)~MeV. (For larger moments the $\alpha_s$ 
expansion~\cite{Broadhurst:1994qj} of the gluon condensate 
contribution~\cite{Shifman:bx} breaks down.)  Using the same 
moments~\cite{Kuhn:2001dm} but choosing $\mu = \hat{m}_q$
instead, we observe a variation of less than 27~(16)~MeV.  This impressive 
improvement clearly overcompensates for the larger $\hat\alpha_s$.
We will choose $\mu = \hat{m}_q$ in the following.  As for the theoretical
uncertainty associated with the truncation of the perturbative series, we
use the method suggested in Ref.~\cite{Erler:1999ug}.
In our case this yields the error estimate,
\be
   \pm N_C Q_q^2 C_F C_A^2 {\hat\alpha_s^3 (\hat{m}_q)\over \pi^3}
       \left( {1\over 2 \hat{m}_q(\hat{m}_q)} \right)^{2n},
\label{error}
\ee
($N_C = C_A = 4 C_F = 3$) corresponding to $\pm 16 \hat\alpha_s^3/\pi^3$ in 
the $\bar{C}_n$. Comparing the corresponding estimate against the exactly known
coefficients of the first eight moments up to order 
$\alpha_s^2$~\cite{Chetyrkin:1996cf,Chetyrkin:1997mb} shows that with
$\mu = \hat{m}_q$, 23 of 24 coefficients are within the estimate, while only 
one coefficient would have been underestimated by a factor 
$\approx 1.437$.  This seems to be a reasonable state of affairs for 
a $1\sigma$ error estimate and corresponds to $\pm 20$~MeV for 
$\hat{m}_c(\hat{m}_c)$ from ${\cal M}_1$, while variation of 
the renormalization scale~\cite{Kuhn:2001dm} assesses this error to only 1~MeV,
which is optimistic.  
We show the estimate~(\ref{error}) in the last column of Table~\ref{moments}.
\begin{table}
\begin{tabular}{|c|ccc|cc|}
n & BES & $\lambda_c^3 = 0.50$ & $\lambda_c^3 = 1.97$ & BES & $\Psi(3S)$ \\
\hline
0 & 5.51      (35) & 5.50 & 7.19 & 0.215      (39) & 0.348      (54) \\
1 & 3.02      (19) & 3.01 & 3.98 & 0.151      (27) & 0.245      (38) \\
2 & 1.68      (11) & 1.68 & 2.25 & 0.106      (19) & 0.172      (27) \\
3 & 0.95 (\five 6) & 0.96 & 1.29 & 0.074      (13) & 0.121      (19) \\
4 & 0.55 (\five 4) & 0.55 & 0.76 & 0.052 (\five 9) & 0.085      (13) \\
5 & 0.32 (\five 2) & 0.33 & 0.45 & 0.037 (\five 6) & 0.060 (\five 9) \\
\end{tabular}
\caption[]{The left part shows contributions to the charm moments 
($\times 10^{n+1}\mbox{GeV}^{2n}$) from $2 M_{D^0} \leq \sqrt{s} \leq 4.8$~GeV,
and the right part from $2 M_{D^0} \leq \sqrt{s} \leq 3.83$~GeV.  Following 
Ref.~\cite{Kuhn:2001dm}, we computed the columns labeled BES by subtracting 
from the threshold data on $R(s)$ the average, $\bar{R}$, below threshold. (We
applied corrections for the leading $s$-dependence.) The errors combine 
the statistical and uncorrelated systematic ones of $\bar{R}$ with those in 
the continuum region and with the common systematics ($\leq 3.5\%$) of 
the difference.}
\label{table:BES}
\end{table}

The last two columns of that Table would agree within errors even if we had
chosen significantly smaller variations in $\lambda_3^b$ and especially 
$\lambda_3^c$ ($\Delta\lambda_3^{b,c} = \pm 1.47$ accounts for the error
introduced by our {\em ansatz\/} and is above and beyond the variations induced
by the fit parameters). The reason for our more conservative error is shown in 
Table~\ref{table:BES}.  It shows that Eq.~(\ref{ansatz}) with 
$\lambda_3^c = 0.50$ reproduces the $n$ dependence of the moments computed from
recent data by the BES Collaboration~\cite{Bai:2001ct} remarkably well. 
However, our method favors $\lambda_3^c \approx \lambda_3^b \approx 1.97$,
and thus 30 to 40\% larger contributions.
Note that the quark parton model predicts $\lambda^q_3 = 1$, while from third
order massive QCD corrections~\cite{Chetyrkin:1996ia} one expects 
$\lambda^q_3 > 1$ (in agreement with our results). 
Table~\ref{table:BES} also compares the BES data to the $\Psi(3S)$ 
contribution~\cite{Hagiwara:2002} in the narrow width approximation.  Even 
assuming that the $\Psi(3S)$ resonance ($M_{\Psi(3S)} = 3.7699$~GeV) 
saturates the charm cross-section in that region, we observe a direct 
{\em experimental\/} $2\sigma$ discrepancy between Ref.~\cite{Bai:2001ct} 
and $\Gamma^e_{\Psi(3S)} = 0.26\pm 0.04$~keV~\cite{Hagiwara:2002}. 
Thus there is a discrepancy between perturbative QCD (our set of sum rules) 
and the BES data, while QCD appears to be consistent with the $\Psi(3S)$ 
data within $1\sigma$. 
The BES data seem to be lower than theoretical predictions by $30 \%$
consistently across the moments.
This constitutes a great puzzle which needs 
to be resolved in the future. We may be able to quote smaller errors after 
this situation has been resolved.
Nevertheless, the quark masses can still be determined precisely enough through
the sum rule approach.

There is a possible contribution from the gluon condensate~\cite{Shifman:bx}.
It is known up to ${\cal O}(\alpha_s)$~\cite{Broadhurst:1994qj}, but its actual
value is not well known.  Its inclusion lowers the extracted quark masses, 
increases $\lambda_3^c$, and sharpens the discrepancy with the BES data.
We can bound its value to $\lsim 0.07\mbox{ GeV}^4$ by demanding $n$ 
independent results within the uncertainties. We use this bound (with a central
value of zero) to account collectively for non-perturbative uncertainties. 
They induce errors of about 29~MeV into $\hat{m}_c(\hat{m}_c)$ ($n = 2$) and 
2.4~MeV into $\hat{m}_b(\hat{m}_b)$ ($n = 6$).

The parametric uncertainties from $\alpha_s$ and the quark masses themselves
are correlated in a complicated way (i) across the moments, (ii) across the two
quark flavors, (iii) between the theoretical moments and the continuum 
contribution, and (iv) with each other.  In practice, all this is accounted 
for by performing fits to the moments.  Heavy quark radiation by light 
quarks~\cite{Portoles:2002rt} is not resonating and problems associated with
singlet contributions~\cite{Portoles:2002rt,Groote:2001py} appear only at 
${\cal O} (\alpha_s^3)$, so these issues should not introduce further 
uncertainties into our analysis.  We will present our final results 
after discussing the $\tau$ lifetime.

\section{$\tau$ lifetime}
It was pointed out long ago that the total hadronic decay width of the 
$\tau$ lepton can be reliably computed in the framework of perturbative 
QCD~\cite{Braaten:1988}. Employing the ratio
\be
R_\tau  =  \frac{\Gamma(\tau \rightarrow \nu_\tau + {\rm Hadrons})} 
  {\Gamma(\tau \rightarrow \nu_\tau e \nu_e)}
\ee
which is predicted to be $N_c=3$ in the lowest order and in the absence of 
Cabbibo mixing, the perturbative corrections can be obtained from QCD 
calculations of the two point current-current correlator
$\Pi_{\mu\nu}(q)$ ($i.e.$, the hadronic contributions to the $W$ boson vacuum 
polarization tensor).  Electroweak radiative corrections were calculated in
Ref.~\cite{Marciano:vm}. Non-perturbative effects have also been estimated and 
found to be small~\cite{Braaten:1991qm}, thus inducing little uncertainty. 
Therefore $R_\tau$ with its small experimental uncertainty provides a solid 
venue to determine the strong coupling constant precisely.

$R_\tau^{\rm QCD}$ can be expressed as a contour integral \cite{Braaten:1991qm}
along $|s|=m_\tau^2$ in the complex $s$-plane,
\ben
R_\tau^{\rm QCD} &=& {1\over2 \pi i } \oint_{|s|= m_\tau^2} 
   \left[ 1 - 2 {s \over m_\tau^2} 
   + 2 \left( {s \over m_\tau^2} \right)^3 - 
   \left( {s \over m_\tau^2} \right)^4 \right] s {d \over ds} \Pi(s),
\label{rtau}
\een
where the Adler function has been calculated to the third order in
$\alpha_S$~\cite{Chetyrkin:1996ia},
\ben
s {d \over ds} \Pi(s) &=& 1 + {\alpha_S(-s)\over\pi} 
+ K_2 \left[{\alpha_S(-s)\over\pi}\right]^2
+  K_3 \left[{\alpha_S(-s)\over\pi}\right]^3 + ...,
\een
where $K_2 = 1.6398$ and $K_3 = 6.371$, assuming three massless flavors of 
quarks.

One way to evaluate  $R_\tau^{\rm QCD}$ is to expand $\alpha_S(-s)$ 
perturbatively in terms of $\alpha_S(m_\tau)$ with the help of 
the renormalization group equation (RGE) of $\alpha_S$.
However, this results in a series of $\alpha_S(m_\tau)$ with poor convergence.
It proves to be expedient to keep the following contour integrals in Eq. (\ref{rtau}) \cite{LeDiberder:1992te}
\ben
A_n & = & {1\over2 \pi i } \oint_{|s|= m_\tau^2} \left[ 1 - 2 {s \over m_\tau^2}
   + 2 \left( {s \over m_\tau^2} \right)^3 - \left( {s \over m_\tau^2} \right)^4 \right] 
\left[{\alpha_S(-s)\over\pi}\right]^n,
\een
which are complicated functions of $\alpha_S$ 
but well-behaved if numerically integrated with the help of the RGE on the complex plane. In general,
\ben
|A_n| \sim \left[{\alpha_S(m_\tau)\over\pi}\right]^n
\een
which we calculate numerically up to 4-loop order in the $\beta$ function~\cite{vanRitbergen:1997va}.

We have included one-loop electroweak radiative corrections \cite{Marciano:vm}:
\ben
S_{EW} = \left(1 + {\alpha \over \pi} \ln {M_Z^2 \over m_\tau^2} \right) 
\left( 1 + {5 \over 12} {\alpha(m_\tau) \over \pi} \right),
\een
where the large log is resumed by the corresponding RGE~\cite{Erler:2002mv}.
We have also included QED (phase space) corrections~\cite{Nir:1989rm}, 
quark condensate contributions, as well as $c$ quark effects in an expansion 
in $m_\tau^2/4 m_c^2$~\cite{Larin:1994va}. 
In total, the partial width into hadrons with vanishing net strangeness is
$$ 
   \Gamma_\tau^{ud} = {G_F^2 m_\tau^5 |V_{ud}|^2\over 64\pi^3}S_{\rm EW}
   ( 1 + {3 m_\tau^2\over 5 M_W^2} ) \left[  {R_\tau^{\rm QCD} \over 3} +
   {\hat\alpha\over \pi} ({85\over 24} - {\pi^2\over 2}) \right.
$$    
\vspace{-12pt}                 
\be
   \left. - 0.09 {m_u^2 + m_d^2\over m_s^2 - m_d^2} -
   {f_{\pi^\pm}^2\over m_\tau^4} [m^2_{\pi^\pm} (8\pi^2 + 23 \alpha_s^2) 
   - 4 m^2_{K^\pm} \alpha_s^2] \right].
\ee

For our analysis, the experimental input is the $\tau$ lifetime,
\be
   \tau_\tau = {\hbar\over \Gamma_\tau} = \hbar {1 - {\cal B}_S
   \over \Gamma_\tau^e + \Gamma_\tau^\mu + \Gamma_\tau^{ud}} =
   290.96 \pm 0.59 \mbox{ fs},
\label{tautau}
\ee
evaluating the partial widths into leptons, 
$\Gamma_\tau^e + \Gamma_\tau^\mu$, as well as $\Gamma_\tau^{ud}$ 
theoretically.  The relative fraction of decays with 
$\Delta S = -1$, ${\cal B}_S = 0.0286 \pm 0.0009$~\cite{Hagiwara:2002}, 
is based on experimental data, since the value for the strange quark mass, 
$\hat{m}_s (m_\tau)$, is not
well known, and the PQCD expansion, $C^{D=2}_{QCD}$, proportional to $m_s^2$ 
converges poorly and cannot be trusted. $C^{D=2}_{QCD}$ also multiplies
the corresponding $m_{u,d}^2$ terms in $\Gamma_\tau^{ud}$, posing the same but 
numerically less important problem there. We solved it, by relating 
$C^{D=2}_{QCD}$ to the ratio $\Gamma_\tau^{us}|V_{ud}|^2/(\Gamma_\tau^{ud} 
|V_{us}|^2) = 0.896\pm 0.034$~\cite{Hagiwara:2002} (in which to linear
order all universal terms cancel), and find 
$C^{D=2}_{QCD} (m_s^2 - m_d^2) = m_\tau^2 (0.091 \pm 0.046)$.

We computed the world average~(\ref{tautau}) by combining the direct value, 
$\tau_\tau = 290.6\pm 1.1$~fs~\cite{Hagiwara:2002}, with
$\tau_\tau ({\cal B}_e,{\cal B}_\mu) = 291.1 \pm 0.7$~fs derived from 
the leptonic branching ratios ${\cal B}_e = 0.1784(6)$ and 
${\cal B}_\mu = 0.1737(6)$~\cite{Hagiwara:2002} taking into account their 1\% 
correlation. The dominant theoretical error induced by the unknown 
coefficient $d_3 = 0 \pm 77$~\cite{Erler:1999ug} is itself strongly 
$\alpha_s$-dependent, is recalculated in each call within a fit, and induces 
an asymmetric $\alpha_s$ error.  

Other experimental uncertainties arise from~\cite{Hagiwara:2002}
$m_\tau = 1.77699(28)$~GeV, $|V_{ud}| =  0.97485(46)$, and 
${\cal B}_S$. Uncertainties from higher dimensional terms in 
the operator product expansion, OPE, are taken from Ref.~\cite{Barate:1998uf} 
and add up to $\Delta \tau_\tau ({\rm OPE}) = \pm 0.64$~fs.  We assume that 
an uncertainty of the same size is induced by possible OPE breaking 
effects\footnote{It is sometimes speculated that OPE breaking effects could
induce dangerous terms of ${\cal O}(\Lambda_{\rm QCD}^2/m_\tau^2)$.
The absence of numerically significant terms of that type is difficult to prove
with rigor. We stress, however, that the fits to OPE condensate terms of 
Ref.~\cite{Barate:1998uf} should have revealed their presence.}.
The unknown five-loop $\beta$-function coefficient, 
$\beta_4 = 0 \pm 579$~\cite{Erler:1999ug}, contributes mainly to the evolution 
of $\alpha_s (m_\tau)$ to $\alpha_s (M_Z)$ and less to the $A_i$. 
The sub-leading errors listed in this paragraph amount to $\pm 1.2$~fs. 
We find, $\alpha_s (m_\tau) = 0.356_{-0.021}^{+0.027}$ and 
$\alpha_s (M_Z) = 0.1221_{-0.0023}^{+0.0026}$, in excellent agreement with 
$\alpha_s (M_Z) = 0.1200 \pm 0.0028$ from $Z$-decays~\cite{Erler:sa} and most
other recent evaluations of $\tau_\tau$~\cite{Barate:1998uf,Raczka:1994ha}. 

\section{Summary}
We have presented a new QCD sum rule with high sensitivity to the continuum regions of
charm and bottom quark pair production.  Combining this sum rule with existing
ones yields very stable results for the \msbar quark masses,
$\hat{m}_c (\hat{m}_c)$ and $\hat{m}_b (\hat{m}_b)$.
Comparison of our approach with experimental data allows for a robust theoretical error
estimate.
We have also provided a new evaluation of the lifetime of
the $\tau$ lepton, $\tau_\tau$, serving as a strong constraint on $\alpha_s$.
Including $\tau_\tau$, and the $n=2$ and $n=6$ moments for the $c$ and $b$ 
quark, respectively, as constraints in a fit to all data~\cite{Erler:sa}
yields,
\be
\ba{l}
  \alpha_s (M_Z) = 0.1211_{-0.0017}^{+0.0018}, \vspace{4pt} \\
  \hat{m}_c (\hat{m}_c)  = 1.289^{+0.040}_{-0.045}~{\rm GeV}, \vspace{4pt} \\
  \hat{m}_b (\hat{m}_b)  = 4.207^{+0.030}_{-0.031}~{\rm GeV}.
\ea
\ee
These results reduce the error~\cite{Erler:1998sy} in $\alpha(M_Z)$ by 25\%.

Finally, we stress again that there is a discrepancy between the BES 
data on one hand, and QCD and the $\Psi(3S)$ electronic width on the other.
The BES data seem to be lower than theoretical predictions
by $30 \%$ consistently across the moments.
This constitutes a great puzzle. 
We may be able to quote smaller errors after this situation has been resolved.

\begin{acknowledgments}
ML would like to thank the High Energy Accelerator Research Organization (KEK)
in Japan for its kind hospitality and 
he is supported in part by a Fund for Trans-Century Talents and CNSF-90103009.
JE is supported by CONACYT (M\'exico) contract 42026--F and by 
DGAPA--UNAM contract PAPIIT IN112902.
\end{acknowledgments}

\end{document}